\documentclass[onecolumn, 10pt]{IEEEtran}

\usepackage{url}
\usepackage{amsmath}
\usepackage{graphicx}
\usepackage{subfigure}
\usepackage{color}
\usepackage{multirow}
\usepackage{algorithm}
\usepackage{algorithmic}
\usepackage{soul}
\usepackage{todonotes}
\usepackage{epstopdf}
\usepackage{amssymb}

\begin{document}
\pagenumbering{gobble} 

\title{Computer-Aided Arrhythmia Diagnosis by Learning ECG Signal}
\author{ 
	Sai Manoj P. D., and Matthias Wess 
\thanks{Preliminary version of this paper is published in ISCAS'17.} }

\maketitle

\begin{abstract} 
	Electrocardiogram (ECG) is one of the non-invasive and low-risk methods to monitor the condition of the human heart. 
	Any abnormal pattern(s) in the ECG signal is an indicative measure of
	malfunctioning of the heart, termed as arrhythmia.
	Due to the lack of human expertise and
	high probability to misdiagnose, computer-aided
	diagnosis and analysis are preferred. 
	In this paper, we perform arrhythmia detection with an optimized neural network 
	having piecewise linear approximation based activation function 
	to alleviate the complex computations in the traditional 
	activation functions. 
	Further, we propose a self-learning method for arrhythmia detection 
	by learning and analyzing the 
	characteristics (period) of the ECG signal. 
	Self-learning based approach achieves 97.28\% of arrhythmia detection accuracy, 
	and neural network with optimized activation functions achieve an arrhythmia detection accuracy of 99.56\%. 
\end{abstract}

\section{introduction}
\label{sec:introduction} 
Amelioration in the cyberphysical systems (CPS) and miniaturization 
of the systems made it feasible to devise
wearable health monitoring and fitness tracking devices in
the form of wristbands, smart watches and much more. 
These health monitoring devices
capture biosignals, which are non-stationary signals representing the 
activity of the organ(s). 

According to world health organization (WHO) cardiovascular disease (CVD) statistics 2015 \cite{WHO'15},
CVDs are the leading factors that cause human death. 
An Electrocardiogram (ECG) is a low-risk, non-invasive technique to represent the physiological state of
heart. Physicians widely use ECG signals for diagnosing arrhythmias
(deviation in ECG signal characteristics in reference to its previous or 
nominal characteristics is a representation of heart disorders) \cite{Thaler'10}.
An Electrocardiogram (ECG) signal is a biosignal that facilitates tracking of 
cardiovascular (heart) activities over time, represented in the form of  
electrical signals.

\begin{figure}[!t]
	\centering
	\includegraphics[width=0.45\textwidth]{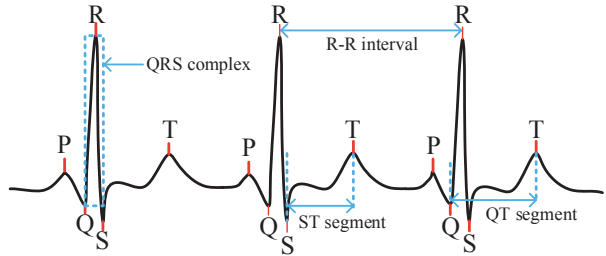}
	\caption{A pseudo-ECG signal and its components}
	\label{fig:ecg_basic}
\end{figure}

In arrhythmia detection, morphology and characteristics of the ECG signal play a vital role.
As seen from Figure \ref{fig:ecg_basic}, ECG signal has different
amplitude levels and shapes marked as P, Q, R, S, and T. 
Different physical activities from individual or group of
heart chambers (arteries, ventricles) generate these components 
\cite{Braunwald'11}. 
An ECG signal is a time series with few millivolts amplitude
and a frequency between 0.01-250 Hz \cite{Webster'10}.
Figure \ref{fig:ecg_basic} shows a pseudo-ECG signal. 
Spatial and temporal properties such as time intervals, the width of individual or
a group of components are employed for arrhythmia detection, such as variation in T component
morphology \cite{Hadjem'15}, ST segment duration, and the
R-R interval \cite{Hadjem'14,Chan'12,Ding'11,Braunwald'11}. 
The terminology used in this paper to refer to ECG signal properties is defined below:

\textit{\textbf{Component}}: It refers to the P, Q, R, S, and T peaks of an ECG signal (Figure \ref{fig:ecg_basic}).

\textit{\textbf{Characteristics}}: It refers to a component's 
inherent properties such as period, amplitude, and width. For instance, R-R
interval is an R peak's characteristic, similarly the width of the QRS complex, or the amplitude of the R peak.~\\

Despite the progress achieved in component detection
and feature extraction in ECG signal, arrhythmia
detection with low computational complexity remains still unanswered.
Feature extraction is referred to the process of obtaining 
the required characteristics of the ECG signal for further processing. 
We outline the primary challenges associated with arrhythmia detection below. 

\subsection{Associated Research Questions in Arrhythmia Detection}
Some of the general challenges associated with arrhythmia detection are:
symptoms of anomalies might not show up all the time;
monitoring the patient and observing ECG for a long time may not be feasible;
and features of the ECG signal vary among different persons and even 
for the same person with time.
As such, there exists no standard metrics that could represent the morphology 
and features of ECG signal that fit all the patients.

To overcome these challenges, computer-aided 
diagnosis (CADiag) is preferred over manual diagnosis to have 
an accurate arrhythmia detection with less false alarms and less human intervention.
The associated research challenges for arrhythmia detection when using CADiag are: 
\begin{itemize}
	\item Optimization of the required amount of data to be processed for accurate arrhythmia detection. 
	\item Machine learning techniques like neural networks though efficient for 
	arrhythmia detection involves computational complexities.  
	\item As there exists no generic set of rules or parameters regarding ECG that can be applied to every one, devising a lightweight arrhythmia detection by learning the characteristics of the ECG signal trading off performance to some extent. 
\end{itemize}
In this paper, we confine to neural networks among the machine learning techniques,
as the arrhythmia detection using neural networks is proven to be one of the efficient techniques for arrhythmia detection \cite{Nambiar'12,Ayub'11}.
However, we compare the arrhythmia detection of proposed methods with other machine learning techniques like the support-vector machine (SVM). 

\subsection{Contributions of this Work}
The above-mentioned research challenges of arrhythmia
diagnosis using CADiag as well the challenges involved in
arrhythmia detection are solved in this paper. 
\textcolor{black}{The primary focus of this work is to detect the arrhythmia rather than determining 
	the type of arrhythmia. 
	The contributions of this work can be outlined in a four-fold manner:
	\begin{itemize}
		\item We propose an optimized neural network architecture preceded with data reduction technique for arrhythmia detection. 
		\item A piece-wise linear approximation for activation
		functions involved in neural networks is proposed to reduce the computational complexity without trading-off performance (accuracy).
		\item An analysis of training algorithms' impact on arrhythmia detection performance is explored and presented. 
		\item Lastly, we introduce a 
		self-learning methodology in which system learns and use the characteristics of the extracted components for arrhythmia detection. 
\end{itemize}}

\section{Literature Review}
\label{sec:lit_revw}

While performing arrhythmia diagnosis, most of the medical experts
take into account the following characteristics: 
the relative positions of the components, magnitudes, morphology, and other
characteristics such as PR interval, PR segment, the width of QRS complex,
QT interval and ST segment \cite{Braunwald'11}. 
There exists various methods for denoising and identifying different components in the 
ECG signal such as Pan-Tompkins \cite{pan'85}.

vast amount of research has been carried out on arrhythmia detection 
and classification.
As the major focus of this work is on arrhythmia detection, we review the 
works on arrhythmia detection that are most relevant to this work. 

For detecting arrhythmias, especially ventricular arrhythmias,
R-R interval values are primarily used \cite{Hadjem'14,Chan'12}.
Based on the variations in the R-R intervals,
statistical parameters such as mean, variance, standard deviation are derived,
and the existence of arrhythmia can be determined
based on the derived statistical parameter values.
Arrhythmia detection is performed with traditional statistical techniques like 
correlation \cite{Bovas'89}, outlier detection \cite{Hadjem'14}, spectrum analysis \cite{Baselli'86} and other
variants as well. 
Advanced predictors and regression techniques 
\cite{Chen'05,Bianco'01} such as auto-regressive integrated moving
average (ARIMA) are also used to predict 
the signal and
compare the predicted and original signals to indicate arrhythmia. 
In \cite{Jekova'00}, autocorrelation functions (ACF953 and ACF994) are
used to analyze the periodicity of the ECG signal, and further utilized to obtain power
spectrum. The arrhythmia detection algorithm performs a linear
regression analysis of ACF peaks. Based on the detected periodicity and the regression
errors, normal sinus rhythms and ventricular fibrillations (VFs) are classified. However,
this technique suffers classifying VFs effectively, 
as VFs might have cosine-like shape. 

Advancements in machine learning techniques led to its adoption in 
multiple applications such as on-chip power and resource management \cite{DAC13,Date_25d'14,ICCAD,ISLPED'14,TCAD13,Slip'15,TC15,TC16,DT16,DT'17,TCAD'18,JOLPE'18,Sayadi'18}, 
hardware security \cite{Jakob_DATE'18,DAC4'18,CF'18,Trustcom'18,CASES'18}, computer vision \cite{CICC15,TCAS'17,Wess_TCAD'18,Sai_ICCAD'18} and other applications. 
Similarly, 
machine 
learning techniques, especially artificial neural networks (ANN) are also 
habituated for arrhythmia detection \cite{Adams'12,Joo'10,Wess'17,GLSVLSI'18}.
Use of artificial neural networks (ANN) for the detection of
left ventricular strain by classifying S-T abnormalities in the ECG 
by Devine and Macfarlane \cite{Devine'93} is one of
the first works that used ANNs for arrhythmia detection.
Several variants of ANN such as like multi-layer perceptron (MLP) \cite{Lim'09,Ceylan'07}, 
modular neural networks (MNN) \cite{Jadhav'10ecg}, general feed-forward neural networks (GFFNN)~\cite{Jadhav'10ecg,Jadhav'10generalized},
radial basis function neural networks (RBFNN) and probabilistic neural networks (PNN)~\cite{Ghongade'14} 
have been implemented for ECG signal analysis and arrhythmia detection.
To befit FPGA's architecture, designers often choose block-based neural networks (BbNNs). 
In general, the BbNNs are trained using evolutionary algorithms such as genetic algorithms
(GA). The Hermite basis function is one of the efficient feature extraction methods for
ECG signals \cite{Chazal'04}. The coefficients of Hermite expansion characterizes
the shape of QRS complex. A BbNN with Hermite expansion coefficients and 
R-R intervals as input is implemented for arrhythmia detection and classification in \cite{Jiang'07}. 
Though neural networks are efficient and effective in detecting arrhythmias, 
most of them are computationally
expensive due to the involved floating point multiplications, 
hyperbolic functions, and other functions.

Other machine learning techniques such as support vector machine (SVM) is also 
employed for arrhythmia detection~\cite{Kohli'10,Nasiri'09}.
For instance, in \cite{Kohli'10}, ECG arrhythmia detection and classification 
using one-against-one (OAO) SVM,
one-against-all (OAA) and fuzzy decision function (FDF) method based SVMs are employed.
In \cite{Nasiri'09}, a genetic algorithm is used
in combination with an SVM classifier for arrhythmia detection.
Other machine learning 
techniques like Bayesian classifiers \cite{Leutheuser'14}, fuzzy logic and 
its variants \cite{Lim'09}, Linear discriminant analysis \cite{Lee'05} 
are as well used for arrhythmia detection. 
Irrespective of the machine learning technique used, in addition to the
involved complex computations and training (for most of the techniques),
the system needs to train every time it has to perform the analysis on a different person.

In this work, we use a neural network 
with resilient propagation (RPROP) with R-R intervals and ECG morphology 
for arrhythmia detection. To reduce the amount of data 
processing, principal component analysis (PCA) 
is performed prior to neural network. This helps to reduce the 
dimensionality of the input data and the computations in the neural network. 
To overcome the computational complexity, the activation functions 
are replaced with piecewise linear approximations.  
Precautionary measures are taken not to compromise the accuracy. In a 
similar vein, we propose a lightweight unsupervised and self-learning 
arrhythmia detection based on the learned ECG characteristics (R-R interval).
Compared to the existing works, the proposed self-learning technique 
requires simple counters and pre-processing 
and does not need to have an explicit training or assumptions about the ECG signal.

\section{Arrhythmia Detection Process}
\label{sec:system}
The proposed workflow for arrhythmia detection along with 
a brief overview of solutions 
given in this section. 

\subsection{Process Overview}

\begin{figure}[!t]
	\centering
	\includegraphics[width=0.45\textwidth]{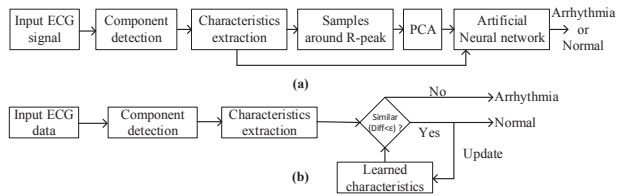}
	\caption{Arrhythmia detection Process: (a) using neural networks; (b) using self-learner}
	\label{fig:system}
\end{figure}

\textcolor{black}{An overview of arrhythmia detection process represented as blocks 
	is 
	presented in Figure \ref{fig:system}. The input ECG data is initially pre-processed, i.e.,
	noise and other artifacts are removed, followed by component detection.
	The component detection and removal of noises are inspired by the work presented 
	in \cite{Pal'10}. Once component detection is performed, the characteristics of the components are extracted. 
	In this work, the R-R intervals and the morphology of ECG signal are used 
	for arrhythmia detection. This is followed by arrhythmia detection process. 
	We propose two different solutions with one based on machine learning 
	(neural networks) and other with self-learning. 
	We represented the processes for two methods in Figure \ref{fig:system} (a) and (b). 
	Neural network based arrhythmia detection (Figure \ref{fig:system}(a)) is accompanied by 
	principal component analysis (PCA) to reduce the dimensionality of the input data. 
	This data will be used by artificial neural network for arrhythmia detection. 
	As the lightweight self-learner (Figure \ref{fig:system}(b)) considers only characteristics of the signal 
	obtained from pre-processing, there is
	no need to implement PCA. Based on the extracted characteristics and the learned characteristics, self-learners signals arrhythmia (if present) and updates the learned characteristics.}

\subsection{Problem Formulation}

For a well established supervised machine learning technique like neural networks which has
demonstrated efficient arrhythmia detection, the computational costs are 
higher due to the involved activation functions. 
As such, the first sub-problem is: ~\\

\emph{\textit{Problem 1: How to optimize the involved computations in a traditional 
		neural network to reduce the computational complexity
		without compromising the performance of arrhythmia detection?}}~\\

Furthermore, as the arrhythmia detection depends on the trained data 
(when using supervised machine learning), it is not accurate or applicable 
when the patient on whom the training is performed and testing is carried out has different heart rhythms. 
As such, the second sub-problem is: 
\newline

\emph{\textit{Problem 2: How to devise an user-independent, unsupervised, yet efficient arrhythmia detection methodology?}} 

\subsection{Overview of Proposed Solutions}
In this work, to solve the problems mentioned above, we propose two solutions. 
\begin{itemize}
	\item To address the complexity involved in the neural network, we propose approximations 
	for the involved activation functions (such as $\tan$, $\tanh$). 
	Further, the data reduction technique such as principal component analysis (PCA) 
	is employed to reduce the number of computations and the size of the neural network. 
	These are performed under the constraints of arrhythmia detection performance and complexity i.e., to to reduce the complexity without trading off (much) performance. 
	\item Secondly, to address the problem of user-independent arrhythmia detection, 
	we propose a self-learning technique where the characteristics of the signal learned by the system, based on which the system determines the signal as arrhythmia or not. 
	For learning the characteristics, simple statistical operations are employed. 
\end{itemize}

\subsection{Performance Metrics}
The parameters used for evaluating arrhythmia detection performance
are accuracy, sensitivity, and specificity. Accuracy (Acc.) is defined as
the number of true positives and true negatives in the total number of samples.
Sensitivity (Sens.) is defined as the ratio of the number of true positives 
to the total number of samples classified as positive (sum of True positives and false negatives). 
Specificity (Spec.) measures the portion of negatives that are correctly identified i.e., the percentage of normal beats identified correctly as normal. The positive 
predictive value (PPV.) represents amount of true positives among all the identified positives. 
The performance metrics can be mathematically expressed as
\begin{equation} \small
\begin{aligned}
\centering
\text{Acc.} & =\frac{TP+TN}{TP+TN+FP+FN}; \qquad 
\text{Sens.}  =\frac{TP}{TP+FN} \\
\text{Spec.} & = \frac{TN}{TN+FP}; \qquad \qquad \qquad \qquad 
\text{PPV}  = \frac{TP}{TP+FP} \\ 
\end{aligned}
\label{eq:metrics}
\end{equation}
\noindent where $TP$, $TN$, $FP$ and $FN$ represents the number of true positives,
true negatives, false positives, and false negatives, respectively.

Accuracy represents the percentage of beats which are correctly identified as normal or arrhythmia; sensitivity represents the percentage of arrhythmia beats correctly identified as arrhythmia; specificity represents the percentage of healthy beats correctly identified as healthy, and positive predictivity represents the percentage of correctly identified arrhythmias among all the beats identified as arrhythmias. Here, the term `correctly' indicates that the indicated condition exists.

\section{Neural Network based Arrhythmia Detection}
\label{trad_nn}
Machine learning is an entrenched domain, employed in the health-care 
field to diagnose different kinds of sicknesses. 
One of the branches of machine learning 
in the arrhythmia diagnosis is the utilization of 
neural networks. 

\textcolor{black}{We first present a fully connected traditional neural network architecture, 
	and activation functions for an effective arrhythmia detection. 
	The utilized traditional neural network comprises of a 
	single hidden 
	layer and an output layer. Neural network architecture with a
	different number of nodes in hidden and output layers 
	are explored, analysis is presented in our previous work \cite{Wess'17}.}

\subsection{Neural Network Architecture and Inputs}
\begin{figure}[!t]
	\centering
	\includegraphics[width=3.4in]{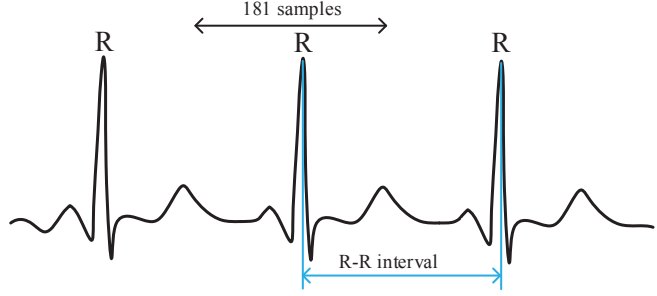} 
	\caption{Input ECG signal used: (a) 181 samples around R component and two R-R intervals; (b) ECG signal inputs to neural network with  PCA}
	\label{fig:ecg}
\end{figure}

To perform arrhythmia detection, a neural network has  
to learn the ECG signal, hence has to be trained with features and/or
morphology of the ECG signal.
In this work, we train the fully connected neural network with 
morphological and temporal features of the
ECG signal. 
The inputs used for the arrhythmia detection are   
two R-R intervals (preceding and succeeding interval) 
to validate the temporal
properties; and 181 samples around the R-peak to learn the morphology of the ECG,
as illustrated in Figure~\ref{fig:ecg}(a).
The input considered for arrhythmia detection comprises of
two R-R interval values, 181 samples around the R peak (183 values in total) 
as the input for processing. 
As designing a neural network with 183 inputs is computationally expensive,
a data reduction technique needs to be employed.

To reduce the dimensionality of the input data, the morphological 
data i.e., 181 values around R peaks are
processed through principal component analysis (PCA).
PCA is a statistical procedure to convert a set of observations of possibly correlated
variables into a set of values of linearly uncorrelated variables. 
The outcome of the PCA is smaller in dimension compared to the size of input data. 
Implementing Fuzzy clustering as in~\cite{Ceylan'07} did not prove
efficient for pre-processing in this 
work, as the major focus is on arrhythmia (anomaly) detection rather than
classification. 
Thus, the PCA has an input of 181 values (only the samples around R peak), 
and the ten principal components from the output of the PCA is
chosen for further processing. 
As such, the input to the neural network is ten values representing the morphological
information and the two R-R intervals, in total 12 values, called as the feature vector.
The feature vector is used as the input for artificial neural network (ANN) to perform 
arrhythmia detection. 
Apart from the architecture and inputs for the neural network, 
the main parameters of choice that affects the performance are the training algorithm 
and the activation functions (in hidden and output layers).

\subsection{Training Algorithm}
There exist multiple learning methods to learn the input data in 
neural networks. As learning has an impact on the performance, 
we first evaluate performances of different learning algorithms 
in the arrhythmia detection application.
Major concerns associated with traditional backpropagation training technique are slow convergence and the
chance for the algorithm to terminate in a local minimum~\cite{Paulin'11}. Paulin et al.~\cite{Paulin'11}
compared the performance of training algorithms for feed-forward artificial neural networks
for classification of breast cancer. It is observed that Levenberg Marquardt (LM) 
performs slightly better
than resilient backpropagation (RPROP)~\cite{Riedmiller'93} and Conjugate Gradient (CG),
in terms of diagnosis accuracy. 
In ~\cite{Kicsi'05} a comparison of RPROP, CG ad LM is performed for 
streamflow forecasting and determination of lateral stress in cohesionless soils.
In the two casestudies, RPROP has higher accuracy than the Levenberg-Marquardt (LM) during the test phase. 
As the existing works are on different kinds of applications and
the performance of training differs with the application, we carry out a case study to
compare the performance of training algorithms.

\begin{figure}[!t]
	\centering
	\includegraphics[width=3.5in]{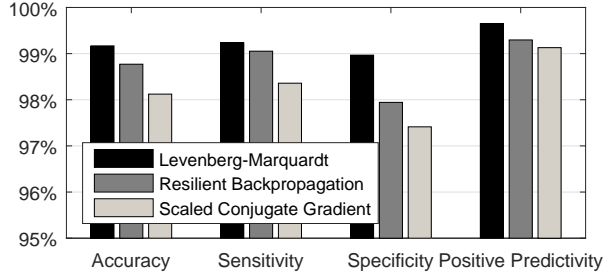} 
	\caption{Performance of different training algorithms in neural network for arrhythmia detection}
	\label{fig:alg_comp}
\end{figure}

Figure \ref{fig:alg_comp} shows the performance of different training algorithms on
ECG signals. Among different algorithms, RPROP is only outperformed by Levenberg-Marquardt algorithm. Considering the complexities \cite{Igel'05} and 
generalization abilities \cite{Paulin'11,Kicsi'05} of the training techniques, 
RPROP was chosen as neural network training algorithm. Though a slight 
degradation is experienced, considering the complexity, we chose RPROP in this work.

\subsection{Activation Function}
For artificial neural networks, several activation functions can be employed, 
such as sigmoid, hyperbolic, Gaussian and so on.
Choosing the best fitting activation function secures the best result for the problem.
In classification problems, the hyperbolic tangent ($\tanh$) function is 
generally used activation function in the hidden layer \cite{Duch'01}.

One of the widely employed activation functions in the output layer 
for the classification problems is the \textit{softmax} function \cite{Duch'01}, 
given by 
\begin{equation}\label{eq:softmax_orig}
\begin{aligned}
\sigma (\mathbf {z} )_{j}={\frac {e^{z_{j}}}{\sum _{k=1}^{K}e^{z_{k}}}} \quad  \forall j = 1,...,K.
\end{aligned}
\end{equation}
where the output vector $\sigma (\mathbf{z})$ is $K$-dimensional; 
the values from the $K$-dimensional input vector $z\in \mathbb{R}^K$ are limited to the range of 0 to 1 and sum up to 1.

\section{Optimized Neural network with PLATanh Functions}
\label{pla_nn}
As the activation functions in the nodes are the most frequently used 
and complex operations in the neural networks, optimizing the activation functions 
without losing accuracy reduces the complexity.

\subsection{Hidden Layer Activation Function Optimization}
\label{subs:opt_hl}

Considering that the final
implementation with fixed-point arithmetic which fits better for resource constrained systems \cite{Wess'17}, the piecewise linear approximation
was performed with gradients to simplify the complex multiplications to simple shift operations~\cite{Hikawa'03}. In comparison to ReLU activation function, $\tanh$ has the advantage of a bounded output making it especially fit for fixed point implementation. 
Some of the previous works ~\cite{Amin'97},~\cite{Hikawa'03} use a maximum of seven ranges. To minimize the error between approximated and exact values from the activation function. In this work, we propose the piecewise linearly approximated hyperbolic
tangent function (PLAtanh) with thirteen ranges. 
The symmetric function is given in (\ref{eq:pla_tanh}), with the limits
given in (\ref{eq:pla_borders}).

\subsection{Output Layer}
The softmax is replaced with the hyperbolic tangent function ($\tanh$). 
While softmax ensures that the sum of the output neurons equals 1, 
with $\tanh$ as the output function, it is possible that none of the 
output values exceed the threshold resulting in no categorization i.e., no output 
provided. This problem is fixed by choosing the highest output value for 
class selection.
In our case, 
to normalize the output layer results between 0 and 1, 
the hyperbolic tangent function is 
\begin{equation}\label{eq:softmax}
Ntanh(x) = \frac{tanh(x)+1}{2}
\end{equation}
An implementation in this form not only allows to also make use of the already
implemented simplified hyperbolic tangent function but also reduces complexity.

Replacing the softmax function in (\ref{eq:softmax_orig}) with Ntanh, given in (\ref{eq:softmax}), as
activation function for output layer neurons, leads to faster convergence
during the training phase and to better fitting due to the small number of
output layer neurons. This difference occurs because in backpropagation
the influence of all output signals on the results of the softmax function
increases the complexity of the algorithm and was therefore simplified in our application. The $\tanh$ in (\ref{eq:softmax}) can be further optimized by the PLAtanh 
given in (\ref{eq:pla_tanh}).

\begin{equation}\label{eq:pla_tanh}
PLAtanh(x) = \left\{
\begin{array}{ll}
1 & x\geq a \\
x/4096+0.9986377 & a\geq x> b \\
x/32+0.905 & b\geq x> c \\
x/8+0.715625 & c\geq x> d \\
x/4+0.53125 & d\geq x> e \\
x/2+0.25 & e\geq x> f \\
x & f\geq x> g \\
x/2-0.25 & g\geq x> h \\
x/4-0.53125 & h\geq x> i \\
x/8-0.715625 & i\geq x> j \\
x/32-0.905 & j\geq x> k \\
x/4096-0.9986377 & k\geq x> l \\
-1 & l\geq x \\
\end{array}
\right.
\end{equation} 
\begin{equation}\label{eq:pla_borders}
\begin{array}{l}
a= 5.5799959, b= 3.02, 
c= 2.02,  
d= 1.475,\\
e= 1.125,  
f= 0.5,  
g= -0.5,  
h= -1.125, \\
i= -1.475, 
j= -2.02, 
k= -3.02,  
l= -5.5799959
\end{array}
\end{equation}
\textcolor{black}{To reduce the computational complexity, the slopes of PLAtanh are chosen as multiples of two. The offsets are chosen to minimize the maximal error of the function, which occurs at $x = 0.5$ with approximately 0.03788.}

\section{Self-Learning based ECG Arrhythmia Detection}
\label{sec:add_hard}
Despite the fact that neural networks are 
efficient in arrhythmia detection, it has to be trained (mostly) and has computational 
complexity though approximations are carried-out. 
We propose an user-independent arrhythmia detection 
technique based on the extracted characteristics of the ECG. 
This method comprises of two steps namely extraction of the signal characteristics 
and arrhythmia detection based on signal characteristics. 

\subsection{Extraction of Signal Characteristics}
\label{subs:feature}
The pre-processing of the ECG signal is carried out with 
discrete wavelet transform (DWT) as described 
in \cite{Pal'10} to detect P, Q, R, S, and T components. 

The identified components play a vital role to extract the characteristics of the 
ECG signal and further processing. 
In this work, the characteristics of interest are the
time interval and the period of detected components. For instance,
if we consider the R component, the characteristics of interest are the interval between
succeeding and preceding R components. 
This extraction can be performed using simple counters on hardware. 
As such, a sample characteristic set is: 
\begin{equation}
\begin{aligned}
C(D)= \{t_{PP},t_{QQ},t_{RR},..., t_{QRS}\}
\end{aligned}.
\end{equation}
\noindent Here $C(D)$ represents the extracted features of the components;
$t_{RR}$ indicates the R-R interval in the ECG signal, and the same goes for Q-Q, P-P and QRS components.
As such, the components are detected and the features are extracted without any
assumptions or explicit training here, thus making it user independent.

The extracted characteristics are 
utilized for arrhythmia detection. 
We consider $t_{RR}$ as the characteristic to determine arrhythmia 
in this work.
The system monitors the characteristics of the components for four consecutive
cycles (four consecutive R-R intervals), and if there is 
no substantial deviation, an average is considered as the characteristic 
that will be used for reference.
The rationale to consider four continuous cycles is that in ECG signals,
most of the irregular heartbeats (arrhythmia) can be
detected by observing four consecutive cycles \cite{Braunwald'11}.

\subsection{Signaling Arrhythmia}
\label{subs:arrhythmia}
To perform the arrhythmia detection, the self-learner makes use of the extracted characteristics (R-R intervals) from
the signal. 
The arrhythmia detection block
validates the characteristics of the incoming signal with the previous learned and stored 
characteristics. 
Unlike machine generated signals, biosignals
will have some variances in the period, amplitudes. For instance, 
a peak can occur
at 0.70s, followed after 0.69s and 0.72s. Hence, tolerance is incorporated. 

When the characteristic (time interval here) of a component is 
similar to the previously observed and learned characteristics, 
the system does not signal anomaly and updates the previously learned characteristics (as in Line \ref{updt1} of Algorithm \ref{algo}). However, when the learned characteristics vary significantly from the 
current characteristic value of the incoming ECG signal i.e., they both differ 
more than a tolerance value as in (\ref{eq:arr}), 
the self-learner signals arrhythmia.

With the R-R feature, the arrhythmia detection output
can be mathematically given as
\begin{equation}
\begin{aligned}\label{eq:arr}
\textnormal{Arrhythmia}=
\begin{cases}
0, & |St_{RR}-t_{RR}|<\varepsilon \\
1, & |St_{RR}-t_{RR}|>\varepsilon
\end{cases}
\end{aligned}
\end{equation}

\noindent \textcolor{black}{where $St_{RR}$ represents the R-R interval learned previously and $t_{RR}$ indicates the
	R-R interval of the beat under analysis. The tolerance is given by $\varepsilon$. 
	The tolerance value is determined by experimentation.} 
The assumption of this process is that the anomalies or arrhythmias do not
occur at the beginning of the process. In case of
anomalies or arrhythmias in the first few samples, i.e., at the beginning of
the setup, the intervals post anomaly will be used as the 
characteristic,
i.e., there will be a slight delay in extracting and learning the characteristics.

\textcolor{black}{ 
	The stored characteristics ($St_{RR}$) are updated regularly in order to keep track 
	of the signal changes and adapt to the user's ECG characteristics rather than using standard values all the time. 
	This provides the advantage of utilizing the derived characteristics from the user rather than 
	using standard values, which might not be valid for all the users. 
	If the characteristics of the incoming
	signal ($t_{RR}$) 
	is similar to the characteristic of the 
	stored beat ($St_{RR}$), then an averaging is performed and the stored
	characteristic ($St_{RR}$) is updated (Line \ref{updt2} of Algorithm \ref{algo}). 
	Thus, the characteristics of the signal used for comparison are updated 
	continuously by learning the signal. This helps in adapting to the 
	signal characteristics of the user.} 

Significant variations in the width or any distortion(s) in the morphology
of an individual component(s) is dealt in the pre-processing stage. Any distortion in 
morphology leads to false detection or not detecting the corresponding 
component, resulting in deriving wrong characteristics which eventually leads to 
indicating arrhythmia at the output. As the characteristics of the signal are 
learnt based on the input ECG signal, this technique is not limited to any single 
individual and can be adapted easily. 
Thus, the user independent and ECG signal based characteristic extraction is performed 
without any explicit training.

\begin{algorithm}[htb]
	\caption{Self-learning based arrhythmia detection}
	\label{algo}
	\begin{algorithmic}[1]
		\REQUIRE ECG signal from sensor ($I(x_1,x_2,...,x_n)$) 
		\ENSURE Arrhythmia detected or do not signal anomaly
		\STATE \textbf{Component Detection:} using Daubechies wavelet as in (\cite{Pal'10})  \label{prep1}
		\STATE QRS complex detection based on DWT coefficients 
		\STATE R component detection with thresholding \label{prep2} 
		\STATE \textbf{Characteristic Extraction} using counters, as in Section \ref{subs:feature} \label{extra1}
		\FOR{Incoming ECG signal} \label{extra2} 
		\IF {$|St_{RR}-t_{RR}|< \varepsilon$} \label{varep1}
		\STATE anomaly $\leftarrow$ `0'\label{varep2}
		\STATE \textbf{update();}\label{varep3}
		\ELSE\label{varep4}
		\STATE anomaly $\leftarrow$ `1'\label{vare5}
		\ENDIF \label{varep6}
		\ENDFOR
		\STATE \textbf{update()}\{  \label{updt1} 
		\STATE \hspace{5mm} $St_{RR}$ $\leftarrow$  $(St_{RR}+t_{RR})/2$ \} \label{updt2}
	\end{algorithmic}
\end{algorithm}

\subsection{Summary:}
A summary of arrhythmia detection process with self-learning is given in 
Algorithm \ref{algo}.
The input ECG signal is initially pre-processed, followed by component detection,
as in Line \ref{prep1}-\ref{prep2}. Further, the characteristics are extracted for the
detected components, as in Line \ref{extra1}-\ref{extra2}.
The process of arrhythmia detection based on the extracted characteristics is given in Line \ref{varep1}-\ref{varep6}.
Here, $St_{RR}$ represents the learned characteristic of the signal and $t_{RR}$ denotes the
characteristic of the signal or beat which needs to be tested for arrhythmia. 
The process of updating the characteristics is given in Line \ref{updt1}-\ref{updt2}.

\section{Simulation Results}
\label{sim_results}
\subsection{Experimental Setup}
An artificial neural network (ANN) with twelve input neurons, one hidden layer 
with six neurons (nodes) is implemented in Vivado HLS tool. The rationale for this 
configuration is the achieved arrhythmia detection accuracy, as 
given in Section \ref{subss:det_eff}. Further, the optimized version of 
the ANN i.e., with PCA,  
PLAtanh as activation function and Ntanh for the output layer has been 
implemented as well in the Vivado HLS. 
The self-learning based arrhythmia detection is implemented using RTL 
in Xilinx ISE as well as in the Matlab. 
As the aim of this paper is to prove the performance of the techniques with optimization,
rather than the hardware resource analysis, we confine to the performance analysis 
in this work. 
For verification of the algorithms, the MIT-BIH arrhythmia database~\cite{PhysioNet} was used.
The database contains forty-eight 30 minute ambulatory ECG recordings,
which also includes unusual but clinically significant arrhythmias.
The database is therefore suitable to evaluate the performance and accuracy of the
optimization and self-learning techniques developed for a wide spectrum of heart diseases~\cite{Moody'01}.
For classification, a variety of feature vectors have been compared in terms
of best classification accuracy.

\subsection{Arrhythmia Detection with Optimized Neural Network}
\begin{figure}[!t]
	\centering
	\includegraphics[width=0.47\textwidth]{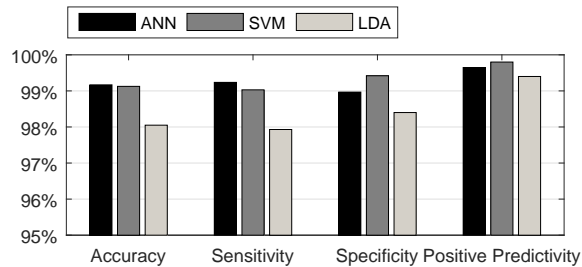} \\
	\caption{Comparison of different machine learning techniques}
	\label{fig:comp_tec}
\end{figure}

We evaluate the performance of different machine learning algorithms for arrhythmia detection first, followed by evaluation 
of actual neural network against neural network optimized activation functions. Figure \ref{fig:comp_tec} shows the 
comparison of arrhythmia detection with neural networks, 
Support vector machines (SVMs) and linear discriminant analysis (LDA). 
\textcolor{black}{It can be seen that the neural networks and SVM perform better than LDA, 
	While in terms of accuracy ANN and SVM achieve the similar performance, their performances differ ($<$0.5\%) in terms of sensitivity specificity and positive predictivity.
	These differences occur due to the class-imbalance of some of the ECG records.
	In terms of computations, SVMs have a higher complexity. 
	Furthermore, performing the approximation of activation functions in neural 
	networks greatly reduces the computational complexity. 
	Additionally, computations in a neural network can be processed in parallel.} 

\subsubsection{Efficiency of Optimized Activation Function}
\begin{figure}[!t]
	\centering
	\includegraphics[width=0.45\textwidth]{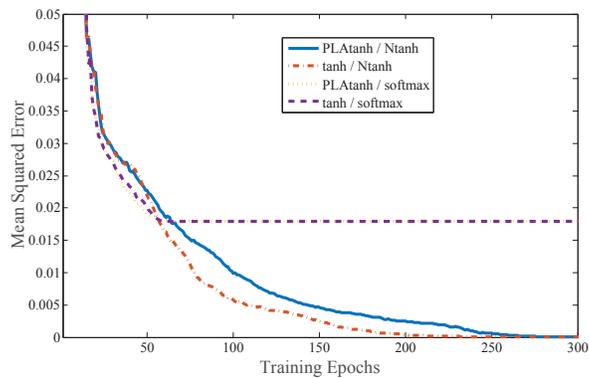} 
	\caption{Mean square error for training with different activation functions}
	\label{fig:activationepochs}
\end{figure}

To evaluate the performance of the activation function (given in (\ref{eq:pla_tanh})) in comparison to the exact
hyperbolic tangent function, neural network is trained once with exact
functions and once with piecewise linear approximations, in both hidden and output
layers. As a typical example, Figure~\ref{fig:activationepochs} shows the mean
square error for one specific record in the database for different activation
functions. The error was logged after every training iteration with RPROP algorithm.
The figure shows that using the piecewise linear approximated activation
function for hidden layer neurons, does not worsen the results in comparison
to exact implementation. 

\begin{figure}[!t]
	\centering
	\includegraphics[width=0.45\textwidth]{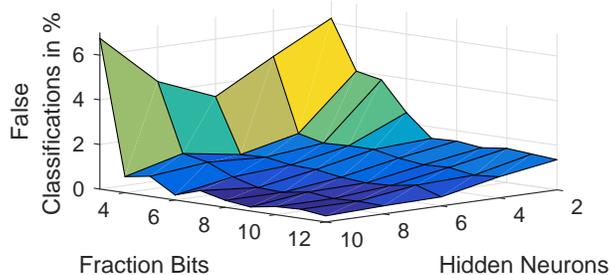} 
	\caption{False classification with different number of hidden neurons and fraction bits}
	\label{fig:fractionhidden}
\end{figure}

\subsubsection{Arrhythmia Detection Efficiency}
\label{subss:det_eff}
\textcolor{black}{We first evaluate the arrhythmia detection performance (in terms of false classification) by varying the number of fraction bits and the number 
	of neurons in the hidden layer. 
	Figure \ref{fig:fractionhidden} shows the false classification error 
	in correlation with the number of fraction bits and number of neurons in the hidden layer. 
	One can observe that the maximum false classification with optimized
	activation functions and a varied number of neurons and fraction bits is nearly 6\%. 
	For better performance, we choose 12 fraction bits and 6 neurons in the hidden layer 
	which has nearly no false classification. Similarly, we evaluated the 
	false classification by varying the number of input neurons in our previous work \cite{Wess'17}. For twelve input neurons and six neurons in hidden layer, 
	the false classification is found to be $<$1\%. Hence, we chose twelve neurons at 
	the input layer and six neurons in hidden layer with 12 fraction bits.} 

With 24 bit data size (12 fraction bits) and other settings described above, 
the optimized neural network approach achieves 99.56\% accuracy.
The achieved performance metrics with optimized and exact implementation of 
neural network is listed in Table \ref{tbl:pparameters}. 
In comparison with exact implementation of neural network, neural network with proposed 
optimized activation functions has nearly 0.03\% degradation in performance. 
This is the trade-off with optimization. \textcolor{black}{Approximation method such as ReLU has 
	slightly lower performance compared to proposed approximation. Additionally, 
	during the experiments it is found that ReLU based implementation has higher false 
	positives and false negatives with same neural network structure than the proposed approximation based implementation.}

\begin{table}[!ht]
	\begin{footnotesize}
		\begin{center}
			\caption{\textcolor{black}{Performance parameters for classification with PLAtanh/Ntanh, tanh/softmax (normal) and ReLU}}
			\label{tbl:pparameters}
			\scalebox{0.9}{
				\begin{tabular}{|c|c|c|c|}
					\hline 
					& \textbf{ PLAtanh } & \textbf{Normal ($\tanh$)} & \textbf{ReLU}  \tabularnewline
					\hline 
					Accuracy & 99.56\% & 99.59\% & 99.15\%\tabularnewline
					\hline 
					Specificity & 99.76\% & 99.79\% & 99.71\%\tabularnewline
					\hline 
					Sensitivity & 99.07\% & 98.06\% & 96.96\%\tabularnewline
					\hline 
					Positive Predictivity & 99.39\% & 99.48\% & 98.88\%\tabularnewline
					\hline 
				\end{tabular}
			}
		\end{center}
	\end{footnotesize}
\end{table}

\subsection{Arrhythmia Detection with Self-learning}
Component detection, including denoising is performed using discrete wavelet transform, 
similar to \cite{Pal'10} in Matlab.  
\textcolor{black}{Based on the experiments carried out for the entire database, 
	the tolerance, $\varepsilon$ is set to 15\% of the corresponding feature.
	For instance, if the R-R interval is 120ms, then a tolerance ($\varepsilon$ in (\ref{eq:arr})) is 18ms.} 

\begin{figure}[!t]
	\centering
	\includegraphics[width=0.45\textwidth]{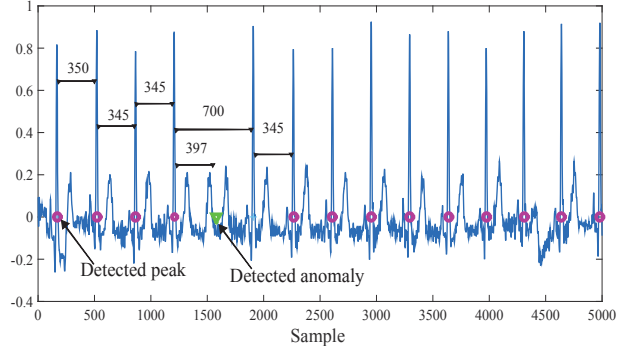} \\
	\caption{Detection of peaks and anomaly in ECG signal}
	\label{fig:anml}
\end{figure}

Arrhythmia detection with self-learning technique in Section \ref{sec:add_hard} 
for an ECG signal with missing R peak is shown in Figure \ref{fig:anml}. 
The time period of the signal is
0.69s (345 samples). From Figure \ref{fig:anml}, we can observe that
first four peaks (circled) are properly detected, because
they are placed at nearly equal time instants. However,
after the fourth peak, the counter waits for the occurrence
of the peak within the time period, and once the period
(+tolerance) is passed and the peak is not encountered,
the circuit outputs an anomaly. The anomaly is indicated
by a (green) triangle marker in Figure \ref{fig:anml}, after 397
samples (period+tolerance). 
Similarly,
the system signals an anomaly if the peak occurs at
a non-periodic time instant.

\begin{table}[!ht] 
	\centering
	\caption{Arrhythmia detection performance of Self-learner}
	\label{tbl:ad_perf}
	\scalebox{0.9}{
		\begin{tabular}{|c|c|c|c|}
			\hline
			Accuracy & Specificity & Sensitivity & Positive Predictivity\\
			\hline
			97.28\% & 98.25\% &  78.70\% & 70.0\% \\
			\hline
		\end{tabular}
	}
\end{table}

The arrhythmia detection performance of the self-learner is given in Table \ref{tbl:ad_perf}.
It has an accuracy of 97.28\%, specificity of 98.25\%, sensitivity of 78.70\% 
and positive predictivity of 70.00\%. 
Sensitivity indicates the percentage of sick people who are correctly identified
as having arrhythmia and specificity indicates the percentage of healthy people
who are identified as healthy.
A low sensitivity and positive predictivity is
achieved due to more number of false negatives and positives resulting from the existence of fusion
beats (which resembles QRS complex), bigeminy and trigeminy ventricular rhythms.

\subsection{Comparison and Discussion}

\begin{table}[!ht]
	\centering
	\caption{\textcolor{black}{Comparison of Arrhythmia detection performance}}
	\label{tbl:comp}
	\scalebox{0.8}{
		\begin{tabular}{|l|c|c|c|l|}
			\hline
			& Acc. & Spec. & Sens. & Note \\
			\hline
			\textbf{ANN} & 99.59\% & 99.79\% & 98.06\% & Full ANN implementation\\
			\hline 
			\textbf{Optimized}  & \multirow{2}{*}{99.56\%} & \multirow{2}{*}{99.76\%} & \multirow{2}{*}{99.07\%} & Optimized ANN \\
			\textbf{ANN} & & & & implementation\\
			\hline 
			\multirow{2}{*}{\textbf{Self-learner}} & \multirow{2}{*}{97.28\%} & \multirow{2}{*}{98.25\%} & \multirow{2}{*}{78.70\%} & Extracted characteristics  \\
			& 			&            &  & based analysis \\
			\hline
			\multirow{2}{*}{\cite{Ahmed'14}} & \multirow{2}{*}{70.00\%} & \multirow{2}{*}{-} & \multirow{2}{*}{-} & One-vs-one error minimization \\
			& & & & with Bayesian classifier\\ 
			\hline
			\cite{Ghongade'14} & 98.10\% & 99.78\% & 98.10\% & Probabilistic ANN \\
			\hline
			\cite{Hamilton'13} & 99.00\% & - & - & Beat-filtering and matching \\
			\hline
			\cite{Adams'12} & 98.60\% & - & - & 3-layer ANN \\
			\hline
			\cite{Jiang'07} &	97.50\% & 98.80\% & 74.90\% & Evolvable Block-based ANN \\
			\hline
			\cite{Ceylan'07} & 93.20-99\% & - & - & FCM-PCA-NN \\
			\hline
			\cite{Hadjem'15} & 92.54\% & 55.4\% & - & Decision tree \\ 
			\hline
			\cite{Nambiar'12} & 99.64\% & & & Block-based ANN \\ 
			\hline
			\cite{Ayub'11} & 99.90\% & - & - & Cascaded ANN \\ 
			\hline 
			\cite{Jadhav'10ecg} & 82.22\% & 82.76\% & 81.35\% & Modular ANN \\ 
			\hline
			\cite{Jadhav'10generalized} & 82.35\% & 89.13\% & 68.18\%& Feed-forward ANN \\ 
			\hline
			\cite{Lim'09} & 97.97\% & 99.20\% & 90.67\% & ANN with fuzzy membership \\ 
			\hline 
		\end{tabular}
	}
\end{table}

\textcolor{black}{Table~\ref{tbl:comp} compares our arrhythmia detection performance with reported performance in other works. 
	Compared to techniques like decision tree, Bayesian classifiers, the neural networks 
	achieves a superior arrhythmia detection performance. We present the comparison of full 
	implementation of feed-forward neural network with the optimized neural network using 
	PLATanh. The optimized neural network 
	performs similar to the full implementation of neural network (ANN). Also, the performance of optimized 
	neural network based implementation is better or similar to other existing works. 
	Compared to $\tanh$,  
	PLAtanh based implementation has nearly 93.5\% reduction in latency, 80.9\% saving in terms of number of 
	LUTs \cite{Wess'17}. The self-learning technique involves much lesser computations (no need 
	of back-propagation, multipliers and so on). 
	The self-learner performs similar to some machine learning techniques in terms 
	of accuracy but has higher false positives, which affects the sensitivity and specificity. The BbNN implementation in \cite{Jiang'07} and proposed
	self-learner has similar performance. 
	The self-learning based method has a slightly lower accuracy ($<$2.5\%) and less 
	sensitivity and positive predictivity in comparison with some neural network implementations.
	This comes at the cost of reduced complexity. 
	Compared to optimized ANN implementation, self-learner utilizes nearly 75\% less number 
	of LUTs and has similar latency.  
	The performance could be further improved by incubating other characteristics such as QRS complex width, period and so on which increases the complexity.}

\textcolor{black}{One can conclude from the above discussion that, neural networks with 
	optimizations are a good fit when computational elements are limited. 
	Under stringent resource constraints where slight performance loss can be tolerated, 
	proposed self-learner is a potential solution.}

\section{Conclusion}
\label{conlusion}
In this work, we have implemented a fully connected neural network for 
arrhythmia detection. To address the complexity constraints in the neural networks, 
piecewise linear approximations for activation functions is proposed. 
Furthermore, a self-learning method that learns the characteristics of the ECG 
signal for arrhythmia detection is devised. The optimized neural 
network achieves an accuracy of 99.56\% and an accuracy with 97.28\% with self-learner 
is achieved for arrhythmia detection. The self-learner suffers from 
false positives which can be improved by learning more characteristics, but at the cost of 
complexity. For much smaller hardware footprints, self-learning method could be adapted
whereas for high accuracy requirements neural networks can be adapted.


\end{document}